\documentclass[letterpaper,12pt]{article}
\usepackage{setspace}
\usepackage{graphicx}
\usepackage{float}
\input{epsf}
\usepackage{txfonts}
\usepackage{subfigure}

\begin{document}
\title{The {\it effective spin} concept\\
to analyze coherent charge transport \\
in mesoscopic systems}
\author{J. Wan, W. Liu, M. Cahay
\\Department of Electrical and Computer Engineering\\
University of Cincinnati, Cincinnati, Ohio 45221\\
\\V. Gasparian\\
Department of Physics, California State University, Bakersfield, CA 93311\\
\\S. Bandyopadhyay \\
Department of Electrical and Computer Engineering \\
Virginia Commonwealth University,
Richmond, Virginia 23284}
\maketitle

\newpage
\vskip 1in

\begin{abstract}
An {\it effective spin} concept is introduced to examine the mathematical and
physical analogy between phase coherent charge transport in  mesoscopic systems and quantum
operations on spin based qubits.  When coupled with the Bloch sphere concept,
this isomorphism allows formulation of transport problems in a language more familiar
to researchers in the field of spintronics and quantum computing.
We exemplify the synergy between charge tunneling and spin
qubit unitary operations by recasting well-known problems
of tunneling through a delta scatterer, a resonant tunneling structure, a superlattice structure,
and arrays of elastic scatterers, in terms of specific unitary operations (rotations) of
a spinor on the Bloch sphere.
\end{abstract}

\parindent      1cm
\vskip .2in
PACS: 73.23.Ad,03.65.Nk,03.67.Lx,76.60.-k

\newpage
\begin{center}
{\bf I. Introduction}
\end{center}

Two major areas of research in condensed matter physics are phase
coherent charge transport in mesoscopic structures
\cite{imry,cahay_review,beenakker} and
spin based quantum computing
\cite{bandy_superlat,bychkov,loss,privman,kane,bandy_prb,all-optical,popescu}.
These two areas are seemingly disparate and until recently have evolved independently.
Some efforts to stress analogies between the two fields
have appeared recently \cite{neder,ionicioiu}.  In this paper, we investigate and develop further the
isomorphism between these two areas by introducing an {\it effective spin} concept to describe phase
coherent charge transport through two-dimensional arrays of elastic scatterers.

\begin{figure}[h]
\centerline{ {\includegraphics[height=4.5in]{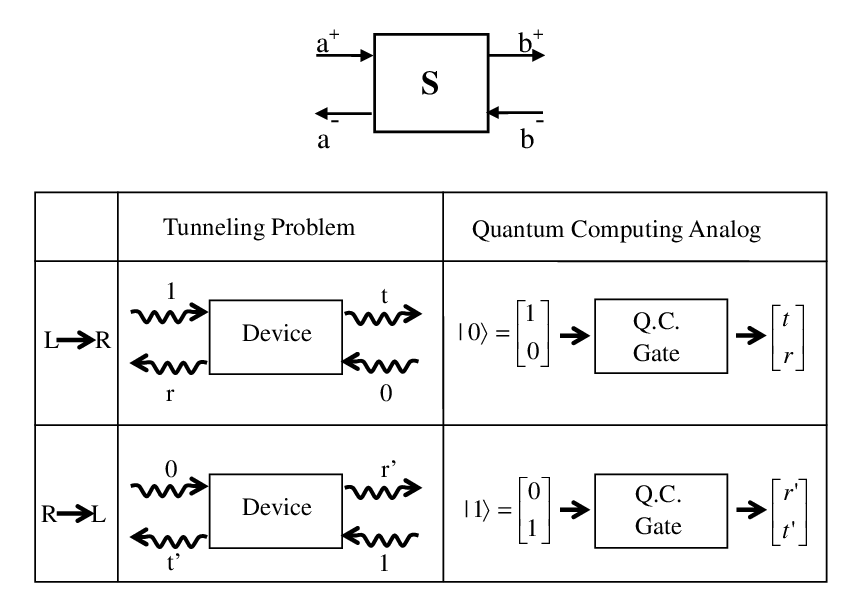}}}
\caption{J. Wan et al} \label{wanfig1}
\end{figure}

In tunneling problems, the mesoscopic structure through which an electron tunnels, is characterized by an arbitrary potential barrier. The transmission and reflection amplitudes are usually calculated by the so-called ``scattering matrix
approach'' \cite{cahay1,cahay2}.
The scattering matrix relates the incoming
$(a^+,b^-)$
to outgoing wave amplitudes $(b^+,a^-)$ on both sides of a
scattering region (mesoscopic structure), as shown in Figure~\ref{wanfig1}, such that
\begin{eqnarray}
|\psi(OUT)\rangle=\left[ \begin{array}{c}
             b^+\\
             a^-
             \end{array}\right]=S\left[ \begin{array}{c}
             a^+\\
             b^-
             \end{array}\right]=\left[ \begin{array}{cc}
             t&r'\\
             r&t'
             \end{array}\right]\left[ \begin{array}{c}
             a^+\\
             b^-
             \end{array}\right]=S|\psi(IN)\rangle,
\end{eqnarray}
where $S$ is the scattering matrix.

For single-mode transport, assuming an electron incident from the left,
\begin{eqnarray}
|\psi^l(IN)\rangle=\left[ \begin{array}{c}
             1\\
             0
             \end{array}\right],
\end{eqnarray}
and
\begin{eqnarray}
|\psi^l(OUT)\rangle=\left[ \begin{array}{c}
             t\\
             r
             \end{array}\right],
\end{eqnarray}
whereas, for an electron incident from the right, we have
\begin{eqnarray}
|\psi^r(IN)\rangle=\left[ \begin{array}{c}
             0\\
             1
             \end{array}\right],
\end{eqnarray}
and
\begin{eqnarray}
|\psi^r(OUT)\rangle=\left[ \begin{array}{c}
             r'\\
             t'
             \end{array}\right].
\end{eqnarray}

The tunneling problem is completely characterized
 by the amplitudes
$(t,r)$ or $(r',t')$ depending on the direction of incidence of the
incoming electron.

Without any loss of generality, we can always think of the two-component column
vector $|\psi(OUT)\rangle$
as a spinor, since it is normalized in the case of
coherent transport. The normalization follows from the unitarity of the
scattering matrix, i.e., $S^\dag{S}=I$. Furthermore, the spinor
$|\psi(OUT)\rangle$ can be thought as the output of
 a one-qubit quantum gate whose input is the spinor
$|\psi(IN)\rangle=(1,0)^\dag$
or $(0,1)^\dag$(where $\dag$ stands for Hermitian conjugate)
depending on the direction of propagation of the incident electron. The
$2\times2$ unitary matrix linking the spinors
$|\psi(IN)\rangle$ and $|\psi(OUT)\rangle$ can therefore be viewed as the
 matrix characterizing rotation of a qubit whose initial state was
$|\psi(IN)\rangle$ and whose final state becomes $|\psi(OUT)\rangle$. This
matrix is also the scattering matrix describing the tunneling problem.
 Herein lies the analogy between quantum logic operation on a
spin qubit and coherent charge transport in a mesoscopic structure.
This paper explores this analogy for single channel charge tunneling through
a single delta-scatterer, a resonant tunneling structure, a periodic array of delta scatterers, and
one-dimensional arrays of randomly distributed elastic scatterers.

\vskip .2in
\begin{center}
{\bf II. Theory}
\end{center}

\begin{figure}[h]
\centerline{ {\includegraphics[height=4.5in]{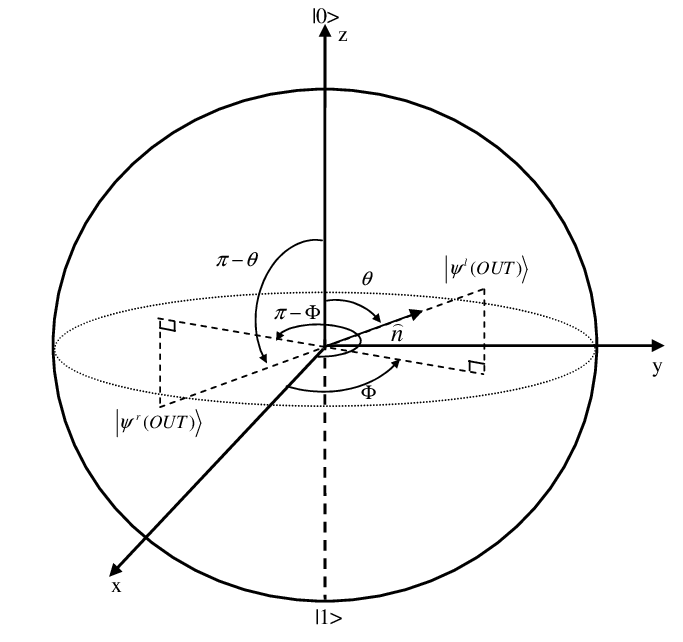}}}
\caption{J. Wan et al} \label{wanfig2}
\end{figure}

Consider the tunneling problem of an electron incident from the
left on an arbitrary one-dimensional conduction band energy profile $E_c(x)$.
We refer to the $(2\times1)$ column vector $|\psi^l(OUT)\rangle$
in Equation (3) as the {\it effective spin} whose components characterize
completely the scattering amplitudes of the tunneling electron. For
an arbitrary potential energy profile $E_c(x)$, the amplitude
$|\psi^l(OUT)\rangle$ can be found by successively  cascading
scattering matrices associated with ``subsections'' within each of which
$E_c(x)$ is
approximated by constant values $E_{c1}$, $E_{c2}$, $E_{c3}$... $E_{cn}$
\cite{cahay1,cahay2}. The evolution of the
pure state $|\psi^l(OUT)\rangle$ after crossing a number of subsections
can be represented using the Bloch sphere concept in which the
spinor is parameterized as follows \cite{nielsen,spinbook}
\begin{eqnarray}
|\psi^l(OUT)\rangle=e^{i\gamma}\left[\cos\frac{\theta}{2}|0\rangle
+\sin\frac{\theta}{2}e^{i\varphi}|1\rangle\right],
\end{eqnarray}
where $\gamma$ is an arbitrary phase factor and the angles
$(\varphi,\theta)$ are the azimuthal and polar angles, as shown in Figure~\ref{wanfig2}.

In Equation (6), $|0\rangle$ and $|1\rangle$ are the $(2\times1)$ column
vectors $(1,0)^\dag$ and $(0,1)^\dag$ respectively, associated with
the north and south poles of the Bloch sphere. They are mutually
orthogonal, i.e., their inner product $<0|1>$ = 0 \cite{nielsen}.

To complete the effective spin picture, we consider
the following $2\times2$ matrix \cite{density}
\begin{eqnarray}
\rho=|\psi^l(OUT)\rangle\langle\psi^l(OUT)|=\left( \begin{array}{c}
             t\\
             r
             \end{array}\right)\left(t^*r^*\right)=\left( \begin{array}{cc
             }
             |t|^2 & tr^*\\
             rt^*  &
              |r|^2
             \end{array}\right).
\end{eqnarray}

Using this density matrix and the Pauli spin matrices ($\sigma_x$,
$\sigma_y $, $\sigma_z $), the effective ``spin components'' associated
with the spinor $|\psi^l(OUT)\rangle$ are given by

\begin{eqnarray}
{\langle}S_x\rangle
=\frac{\hbar}{2}\tt{Tr}\left(\rho\sigma_x\right)
=
\frac{\hbar}{2}\left(tr^*+rt^*\right)={\hbar}\tt{Re}\left(rt
^*\right)={\hbar}\tt{Re}\left(r^*t\right),
\end{eqnarray}

\begin{equation}
{\langle}S_y\rangle
=\frac{\hbar}{2}\tt{Tr}\left(\rho\sigma_y\right)
=\frac{\hbar}{2}i\left(tr^*-rt^*\right)={\hbar}\tt{Im}\left(r
t^*\right)=-{\hbar}\tt{Im}\left(r^*t\right),
\end{equation}
and
\begin{eqnarray}
{\langle}S_z\rangle
=\frac{\hbar}{2}\tt{Tr}\left(\rho\sigma_z\right)
=\frac{\hbar}{2}\left(|t|^2-|r|^2\right)
=\frac{\hbar}{2}\left(1-2|r|^2\right)=\frac{\hbar}{2}\left(2|t|^2-1\right).
\end{eqnarray}

For an electron incident from the right,
$|\psi^r(IN)\rangle=|1\rangle$, and the density matrix \\
$\rho'$ (=$|\psi^r(OUT)\rangle\langle\psi^r(OUT)|$)  is such that $\rho'=1-\rho$, where
$\rho$ is given
by Equation (7) and the components ${\langle}S_x\rangle$,
${\langle}S_y\rangle$ and ${\langle}S_z\rangle$ are just the
negative of the values in Equations (8-10).  Therefore the two spinors
corresponding to $|\psi^l(OUT)\rangle$ and $|\psi^r(OUT)\rangle$ are mirror
images of each other, corresponding to a reflection
through the origin of the Bloch sphere. This means  that $|\psi^l(OUT)\rangle$
and $|\psi^r(OUT)\rangle$ are orthogonal, which they must be because the
scattering matrix is unitary.

The unitarity of the scattering matrix also leads to:
\begin{eqnarray}
{\langle}S_x\rangle^2+{\langle}S_y\rangle^2=\hbar^2|t|^2\left(1-|t|^2\right),
\end{eqnarray}
and
\begin{eqnarray}
{\langle}S_x\rangle^2+{\langle}S_y\rangle^2+{\langle}S_z\rangle^2=\hbar^2/4.
\end{eqnarray}

Equation (11) shows that
the projection of the spinor in the equatorial plane of the Bloch
sphere reaches a maximum when $|t|=|r|=1/\sqrt{2}$.
Actually, ${\langle}S_x\rangle^2+{\langle}S_y\rangle^2$ is
proportional to $|t|^2\left(1-|t|^2\right)$, i.e., the low frequency
shot noise power for the tunneling electron \cite{blanter}.
Since $\left({\langle}S_x\rangle,{\langle}S_y\rangle,{\langle}S_z\rangle\right)$
are proportional to the components of the spinor $|\psi(OUT)\rangle$ on the
Bloch sphere,
Equation (12) simply states that the spinor stays on the Bloch sphere during
cascading of scattering matrices. This is expected for the case of coherent
transport.  The angles $(\gamma,\theta,\varphi)$ appearing in the generic
expression of the spinor (or qubit) in Equation (6) can be expressed in
terms of the phases and magnitudes of the reflection and transmission
coefficients:
\begin{eqnarray}
|\psi^l(OUT)\rangle=\left[\begin{array}{c}
             t\\
             r
             \end{array}\right]=\left[\begin{array}{c}
             |t|e^{i\phi_T}\\
             |r|e^{i\phi_R}
             \end{array}\right]=e^{i\phi_T}\left[\begin{array}{c}
             |t|\\
             |r|e^{i(\phi_R-\phi_T)}
             \end{array}\right].
\end{eqnarray}
where $\phi_R$ and $\phi_T$ are the phases of the reflection and transmission
amplitudes, respectively.

We get
\begin{eqnarray}
\gamma=\phi_T,
\end{eqnarray}
and
\begin{eqnarray}
\varphi=\phi_R-\phi_T.
\end{eqnarray}
Furthermore,
\begin{eqnarray}
|t|=\cos\frac{\theta}{2},
\end{eqnarray}

\begin{eqnarray}
|r|= \sin\frac{\theta}{2} = \sqrt{1-|t|^2},
\end{eqnarray}
and therefore,
\begin{eqnarray}
\frac{\theta}{2}=\tan^{-1}\left(\frac{|r|}{|t|}\right).
\end{eqnarray}

Equations (8-10) are therefore equivalent to
\begin{eqnarray}
{\langle}S_x\rangle = \frac{\hbar}{2}\sin\theta\cos\varphi,
\end{eqnarray}

\begin{eqnarray}
{\langle}S_y\rangle = \frac{\hbar}{2}\sin\theta\sin\varphi,
\end{eqnarray}
and
\begin{eqnarray}
{\langle}S_z\rangle = \frac{\hbar}{2}\cos\theta,
\end{eqnarray}

Equations (8) and
(9) clearly show that the averages ${\langle}S_x\rangle $ and
${\langle}S_y\rangle $ contain more information than the sample conductance
alone. The latter depends only on the magnitude of transmission $|t|$ or
reflection $|r|$ in the Landauer picture \cite{landauer}, whereas
${\langle}S_x\rangle $ and
${\langle}S_y\rangle $ depend on the phase relationship between $t$ and $r$ as
well. The
phase relationship is a strong
function of the energy of the incident electron. At non zero
temperature, there will be a thermal spread in the energy of the incident
electron which will lead to a rapid wash out with
temperature of the components ${\langle}S_x\rangle $ and
${\langle}S_y\rangle $, i.e., the off-diagonal components of the
density matrix $\rho$. Note that while ${\langle}S_x\rangle $ and
${\langle}S_y\rangle $ depend on the off-diagonal components of the density
matrix and are very energy sensitive, ${\langle}S_z\rangle $ depends only on the
diagonal components of the density matrix and is much less energy sensitive.

\vskip .1in
\parindent 0cm
{\bf II.1 Quantum computing gate analog}
\parindent 1cm

The $2\times2$ unitary matrix or quantum computing gate $U_{QG}$ which relates
$|\psi(OUT)\rangle$ and $|\psi(IN)\rangle$ on the Bloch sphere has the general
form \cite{nielsen}

\begin{eqnarray}
U_{QG} (\alpha,\beta,\eta,\zeta)=e^{i\alpha}R_z(\beta)R_y(\eta)R_z(\zeta),
\end{eqnarray}
where $(\alpha,\beta,\eta,\zeta)$ are real numbers and the $R_y$ and
$R_z$ are the $2\times2$ matrices associated with rotations of the spinor on the
Bloch sphere about the $\widehat{y}$ and $\widehat{z}$ axis,
respectively. Using the fact that
$R_y(\eta)=e^{-i\frac{\eta}{2}\sigma_y}$ and
$R_z(\zeta)=e^{-i\frac{\zeta}{2}\sigma_z}$\cite{nielsen}, we obtain:
\begin{eqnarray}
U_{QG} (\alpha,\beta,\eta,\zeta)=\left[\begin{array}{cc}
e^{i \left(\alpha - \frac{\beta}{2}-\frac{\zeta}{2}\right)} \cos\frac{\eta}{2}
& - e^{i \left(\alpha - \frac{\beta}{2}+\frac{\zeta}{2}\right)} \sin\frac{\eta}{2}\\
  e^{i\left(\alpha + \frac{\beta}{2} - \frac{\zeta}{2}\right)}\sin\frac{\eta}{2}
& e^{i\left(\alpha + \frac{\beta}{2} + \frac{\zeta}{2}\right)}\cos\frac{\eta}{2}\\
\end{array}\right].
\end{eqnarray}

For $|\psi^l(IN)\rangle=|0\rangle$, we have
\begin{eqnarray}
|\psi^l(OUT)\rangle = U_{QG} (\alpha,\beta,\eta,\zeta) |0\rangle=\left[\begin{array}{c}
    e^{i\left(\alpha-\frac{\beta}{2}-\frac{\zeta}{2}\right)}\cos\frac{\eta}{2}\\
     e^{i\left(\alpha+\frac{\beta}{2}-\frac{\zeta}{2}\right)}\sin\frac{\eta}{2}
             \end{array}\right],
\end{eqnarray}
which is the special case of a spinor on the Bloch sphere
in Equation (6), corresponding to
\begin{eqnarray}
\alpha & = & \gamma=\phi_T, \nonumber \\
\eta & = & \theta=2tan^{-1}\left [ \frac{ |r|}{|t|} \right ], \nonumber \\
\beta & = & -\zeta=\varphi=\phi_R - \phi_T.
\end{eqnarray}

Hence, from a quantum computing perspective, the analytical expression for $U_{QG}$ is identical
to the scattering matrix used to described the tunneling problem
and is given explicitly by
\begin{eqnarray}
U_{QG} ( {\phi_T}, \theta, |t| ) =
e^{i\phi_T}R_z(\phi_R - \phi_T)R_y
\left(2tan^{-1}\left [ \frac{ |r|}{|t|} \right ] \right)R_z(\phi_T-\phi_R),
\end{eqnarray}
This last equation helps visualizing coherent
charge transport (or tunneling) through specific
mesoscopic devices as a successive set of rotations of the effective spin
on the Bloch sphere, as will be illustrated in the numerical examples in section III.

In the next section, we provide several examples to illustrate the {\it effective
spin} concept.

\bigskip

\begin{center}
{\bf III. Examples}
\end{center}

\vskip .1in
\parindent 0cm
{\bf III.1 Scattering across a single delta scatterer}
\parindent 1cm

We first determine the quantum computing gate analog of a simple
delta scatterer of strength $V_I\delta(x)$ for which the reflection
and transmission amplitudes are easily shown to be
\begin{eqnarray}
t'=t=\frac{ik}{ik-k_0}=\frac{i\widetilde{k}}{i\widetilde{k}-1},
\end{eqnarray}
and
\begin{eqnarray}
r'=r=\frac{k_0}{ik-k_0}=\frac{1}{i\widetilde{k}-1},
\end{eqnarray}
with $\widetilde{k}={k/k_0}$, $k_0=m^\star{V_I}/\hbar^2$ and
$k=\frac{\sqrt{2m^\star{E}}}{\hbar}$, where $E$ is the kinetic
energy of the electron and $m^\star$ is its effective mass.

The magnitude and phase of $t$ and $r$ are therefore
\begin{eqnarray}
|t|=\frac{\widetilde{k}}{\sqrt{\widetilde{k}^2+1}},
\phi_T=-\tan^{-1}\left(\frac{1}{\widetilde{k}}\right),
\end{eqnarray}
and
\begin{eqnarray}
|r|=\frac{1}{\sqrt{\widetilde{k}^2+1}},
\phi_R=\tan^{-1}\left({\widetilde{k}}\right)-\pi.
\end{eqnarray}
The spinor $|\psi^l(OUT)\rangle$ for this simple problem is given by
Equation (6), where
\begin{eqnarray}
\varphi=\phi_R-\phi_T=-\frac{\pi}{2},
\end{eqnarray}
and
\begin{eqnarray}
\theta=2\tan^{-1}\left(1/{\widetilde{k}}\right).
\end{eqnarray}
The equivalent quantum computing gate is characterized by unitary
matrix $U_{QG}$ given by
\begin{eqnarray}
U_{QG}=e^{i\varphi_T}R_z(\frac{-\pi}{2})R_y(\theta)R_{z}(\frac{\pi}{2})
=e^{i\varphi_T}R_x(-\theta),
\end{eqnarray}
where $R_x$ is the matrix for spinor rotation around the x-axis \cite{nielsen}.
For low incident energy, $\theta=\pi$ and it monotonically goes to
$0$ as the energy of the incident electron increases. According to
Eqns.(19-21), the spinor $|\psi^l(OUT)\rangle$ sweeps only a very limited
portion of the Bloch sphere, i.e., the semi-circle in the y-z plane,
going from the south to north poles clockwise as the energy of the
incident electron increases. The spin components of
$|\psi^l(OUT)\rangle$ along the x, y, and z axes are given by
\begin{eqnarray}
\langle{S_x}\rangle=0,
\end{eqnarray}
\begin{eqnarray}
\langle{S_y}\rangle=-\frac{\hbar}{2}\left(\frac{2\widetilde{k}}{\widetilde{k}^2+
1}\right),
\end{eqnarray}
and
\begin{eqnarray}
\langle{S_z}\rangle=\frac{\hbar}{2}\left(\frac{\widetilde{k}^2-1}
{\widetilde{k}^2+1}\right).
\end{eqnarray}

For instance, when $\widetilde{k}=1$, $|\psi^l(OUT)\rangle$ is in the
equatorial plane of the Bloch sphere, along the y-axis. In this
case, $\theta=\frac{\pi}{2}$, and the matrix $U_{QG}$ is given by
\begin{eqnarray}
U_{QG}=e^{-i\frac{\pi}{4}}R_z(-\frac{\pi}{2})R_y(\frac{\pi}{2})R_z(\frac{\pi}{2})
=e^{\frac{-i\pi}{4}} S\left(-\frac{\pi}{2}\right)\sigma_x{\cal
H} S\left(\frac{\pi}{2}\right),
\end{eqnarray}
where
\begin{eqnarray}
S(\delta)=\left[\begin{array}{cc}
            1 & 0\\
            0 & e^{i\delta}
             \end{array}\right],
\end{eqnarray}
and
\begin{eqnarray}
{\cal H}=\frac{1}{\sqrt{2}}\left[\begin{array}{cc}
            1 & 1\\
            1 & -1
             \end{array}\right],
\end{eqnarray}
are the general phase shift and the Hadamard matrix, respectively,
extensively used in the theory of quantum computing \cite{nielsen}.

\vskip .1in
\parindent 0cm
{\bf III.2 Scattering through a delta-scatterer in a region of length $a$}
\parindent 1cm

Next, we consider the scattering problem across a region of length a
containing a delta scatterer at location $x_0$.
The corresponding scattering matrix can be easily derived. The location of
the spinor $|\psi^l(OUT)\rangle$ on the Bloch sphere is described by azimuthal angle
$\theta$ given in Equation (21) and polar angle
\begin{eqnarray}
\varphi=\frac{-\pi}{2}-k\left(a-2x_0\right).
\end{eqnarray}

The average values of the effective spin components are given by

\begin{eqnarray}
\langle{S_x}\rangle & = &
\frac{\hbar}{2}\left(\frac{2\widetilde{k}}{\widetilde{k}^2+1
}\right)\sin{k\left(2x_0-
a\right)}, \\
\langle{S_y}\rangle & = -&
\frac{\hbar}{2}\left(\frac{2\widetilde{k}}{\widetilde{k}^2+1
}\right)\cos{k\left(2x_0-a\right)}, \\
\langle{S_z}\rangle & = & \frac{\hbar}{2}\left(\frac{\widetilde{k}^2-1}
{\widetilde{k}^2+1}\right).
\end{eqnarray}

In this case, $\langle{S_x}\rangle$ is non-zero unless
$x_0=\frac{a}{2}$, i.e., unless  the
potential energy profile in the device is spatially symmetric. For a fixed value
of the
incident wavevector, the spinor $|\psi^l(OUT)\rangle$ moves on a
circle parallel to the $(x,y)$ plane. If $a$ is selected such that
$ka=\pi$, $\varphi$ increases linearly from $-\frac{3\pi}{2}$ to
$\frac{\pi}{2}$ as $x_0$ varies from $0$ to $a$, i.e., the Bloch
vector associated with the spinor sweeps the entire plane
defined by the component $\langle{S_z}\rangle$.
According to Eqns.(41) and (42), if $ka=\pi$, the average value of
$\langle{S_x}\rangle$ and $\langle{S_y}\rangle$
are equal to zero when we average over the impurity location $x_0$.
This is an important ingredient in the theory of localization
in 1D arrays of scatterers, as will be discussed later.
The quantum computing gate $ U_{QG}$ analog of this tunneling problem is
given by
\begin{eqnarray}
U_{QG} = e^{i\phi_T}R_z\left(-\frac{\pi}{2}
+k(2x_0-a)\right)R_y(\theta)R_z\left(k(a-2x_0)+\frac{\pi}{2}\right).
\end{eqnarray}
Since $\theta$ is still given by Equation (21), a {\it spin flip} from the south
to north pole is only possible if we increase the energy of the
incident electron to infinity. The energy cost for the spin flip is
drastically reduced if we have two or more delta scatterers,
as discussed next.

\vskip .1in
\parindent 0cm
{\bf III.3 Scattering across a resonant tunneling structure}
\parindent 1cm

\begin{figure}[h]
\centerline{
{\includegraphics[height=6.5in,angle=270]{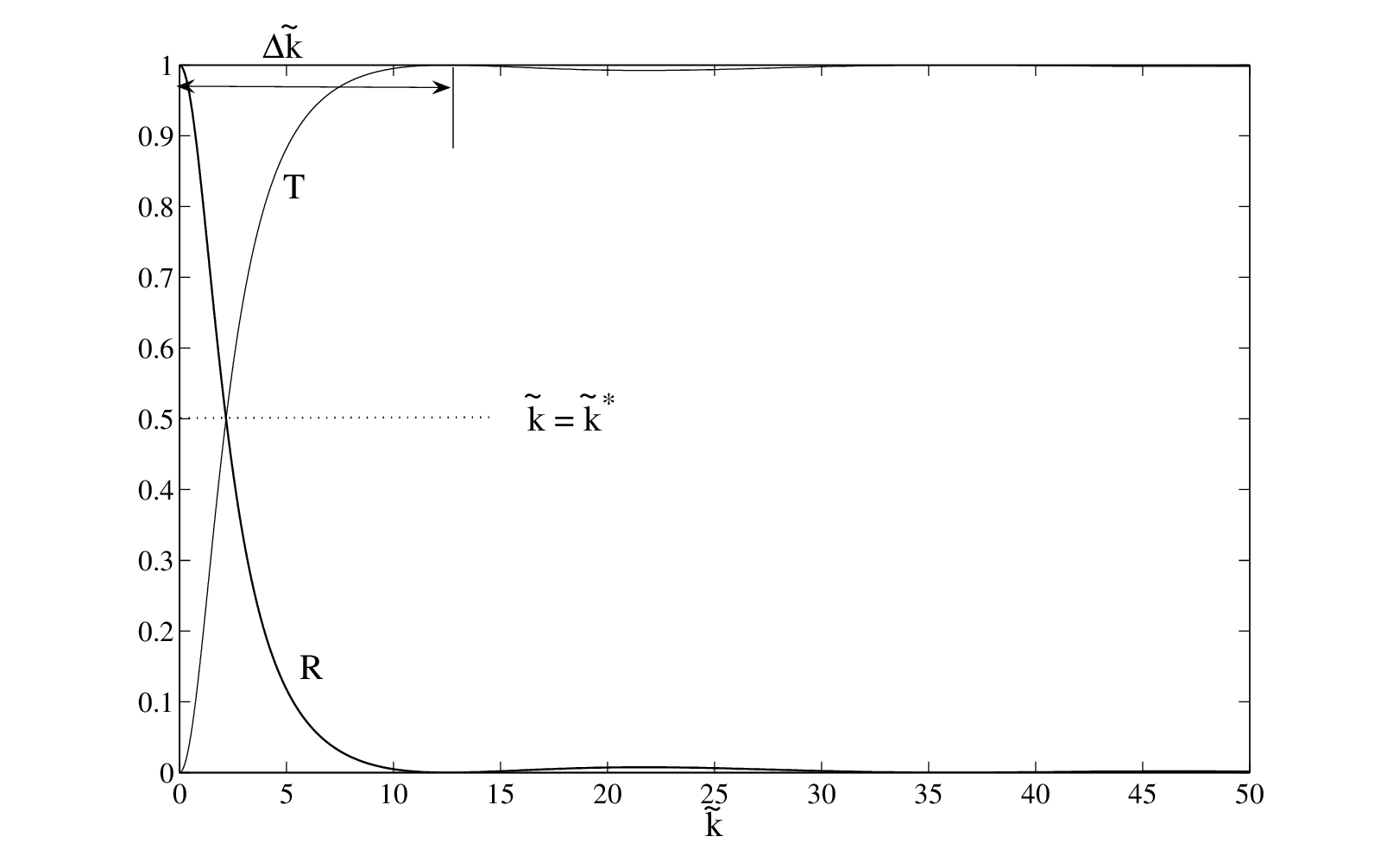}}}
\caption{J. Wan et al}
\label{wanfig4}
\end{figure}

\begin{figure}[h]
\centerline{ {\includegraphics[height=4.3in]{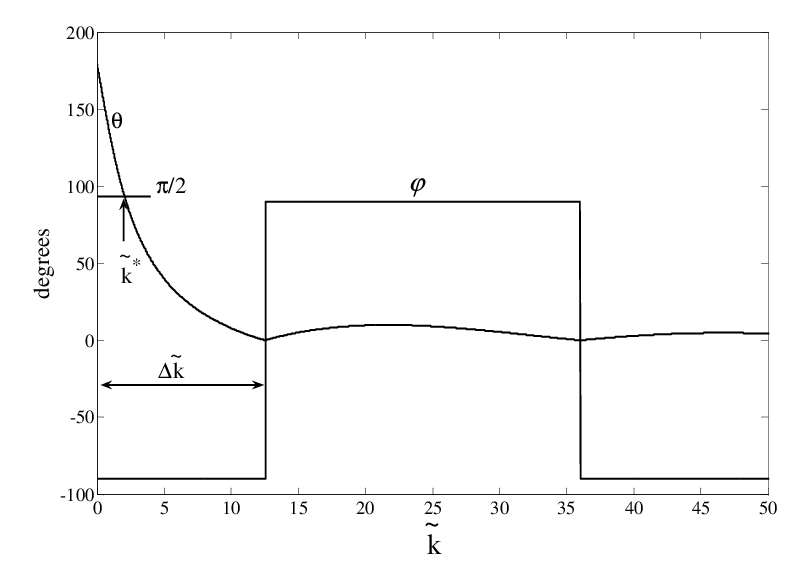}}}
\caption{J. Wan et al} \label{wanfig5}
\end{figure}

We consider the scattering problem across a resonant tunneling structure
consisting of two delta scatterers of
equal strength $V_I$ separated by a distance a.
In our numerical simulations, we use $V_I=0.3eV {\AA}$ and $a$ = 50 ${\AA}$.
Figure~\ref{wanfig4} is a plot of the transmission coefficient $T$ as a function of the
reduced wavevector $\widetilde{k}$. The first two resonances (at
which $T=1$) occur at $\widetilde{k}\approx 12.5$ and $36$. The corresponding
variation of the phase angles $(\varphi,\theta)$ for the spinor
$|\psi^l(OUT)\rangle$ are displayed in Figure~\ref{wanfig5}.
The angle $\theta$ reaches its minimum value of zero at
the resonances when there is a sudden jump in $\varphi$.
When viewed as a quantum computing gate,  an RTD is
more efficient when operated over the range
$\Delta\widetilde{k}$ indicated in Figure~\ref{wanfig5} since it allows a full
swing in $\theta$ from $0$ to $\pi$, whereas the swing in $\theta$
is much smaller between the first two and higher resonances. The quantity
$\Delta\widetilde{k}$ is much smaller than the infinite change in
$\widetilde{k}$ needed for a single delta-scatterer to realize an
inverter, as discussed in the previous section.
Since $T=R$ for $\widetilde{k}=\widetilde{k}^\star$, $\theta=\frac{\pi}{2}$
which is
enough to implement the Hadamard gate using an RTD.

\begin{figure}[h]
\centerline{
{\includegraphics[height=6.5in,angle=270]{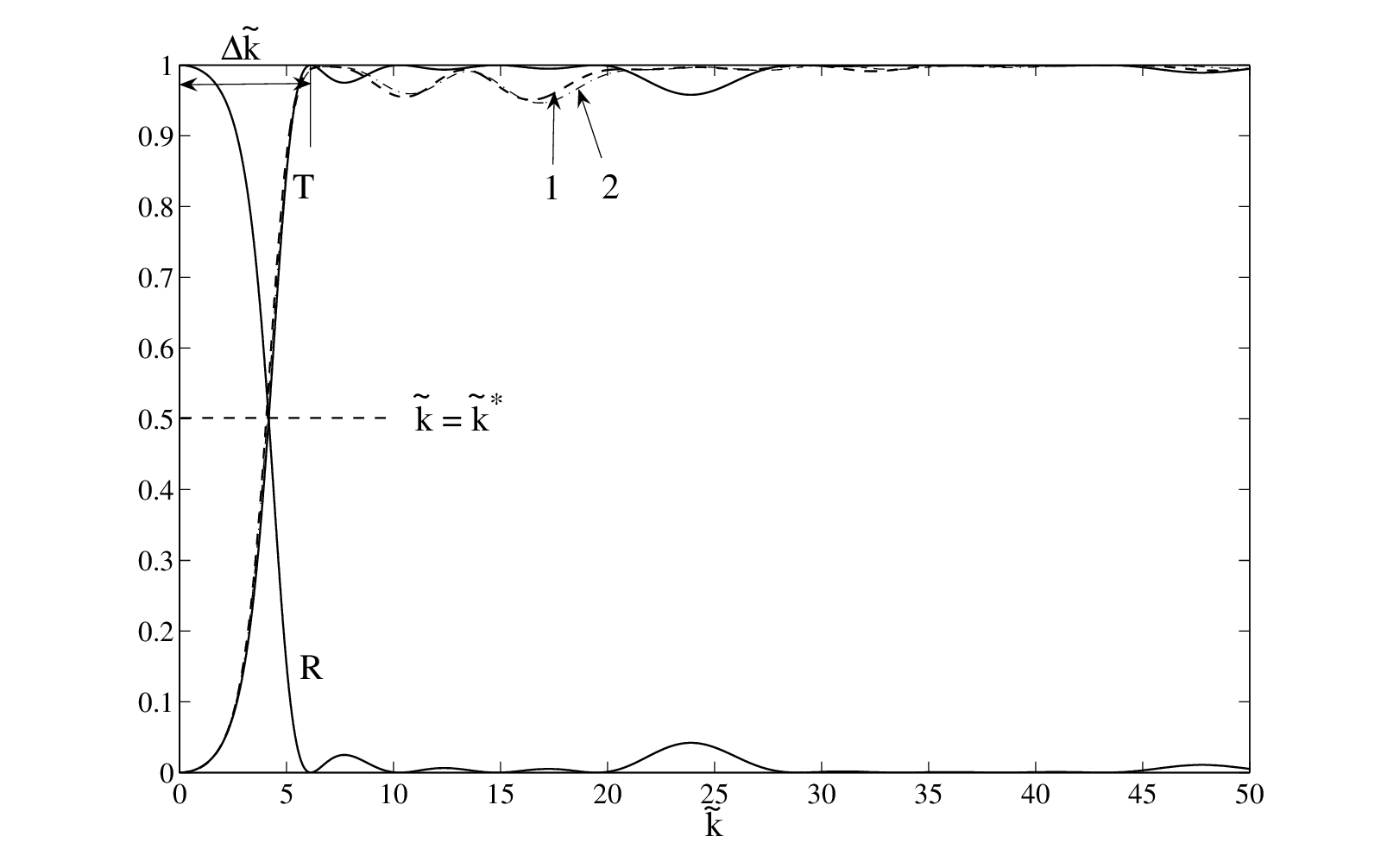}}}
\caption{J. Wan et al}
\label{wanfig6}
\end{figure}

\begin{figure}[h]
\centerline{ {\includegraphics[height=4.3in]{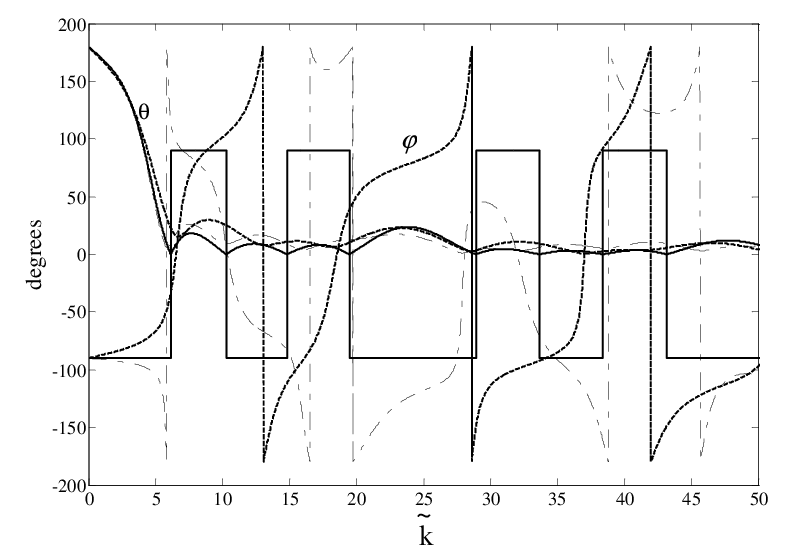}}}
\caption{J. Wan et al} \label{wanfig7}
\end{figure}

The results above can be extended to the case of a superlattice,
modeled as a sequence of evenly spaced identical delta scatterers.
In that case, each resonant state present in the smaller unit with
two scatterers (RTD) leads to a passband for the infinitely periodic
structure. In Figure \ref{wanfig6}, we plot the transmission coefficient for a
structure consisting of 5 delta scatterers with the same parameters
as for the RTD described above and with the same spacing of 50 ${\AA}$
between each scatterer. The transmission coefficient reaches
unity at four values of $\widetilde{k}$ in the interval $[5 - 25]$,
which is a well known result for finite repeated structures
\cite{vezetti,cahay3}. Furthermore, the range $\Delta\widetilde{k}$ needed
to reach the condition $T=R$ is reduced compared to the case of a RTD.
As the number of periods in the superlattice increase,
$\Delta\widetilde{k}$ actually converges to a limit corresponding to
the lower edge of the pass band of the infinite superlattice. As
shown in Figure~\ref{wanfig7}, the angle $\theta$ allows a full swing from north
to south poles on the Bloch sphere over a range $\Delta\widetilde{k}$
smaller than what is necessary for the case of the RTD, and the phase angle
$\varphi$
 toggles back and forth between $-\frac{\pi}{2}$ and
$\frac{\pi}{2}$ each time a resonance is crossed.

Figure \ref{wanfig6} also shows a plot of the transmission
coefficient (curves labeled 1 and 2) versus $\widetilde{k}$ for two
imperfect structures, in which the locations of the five delta
scatterers are selected randomly and uniformly over each interval of
length $a$. The transmission coefficient is fairly sensitive to
$\widetilde{k}$ in the range of $\widetilde{k}$ where the lower pass
band will develop for the infinite superlattice. However, the
transmission curve is fairly insensitive to the imperfections in the
superlattice in the same range of $\widetilde{k}$. As shown in
Figure~\ref{wanfig7}, the angle $\theta$ is also fairly insensitive
to imperfections in the superlattice but the phase $\varphi$ is not.
The latter result is a compounded effect of multiple reflections
between impurities and the sensitivity  of $\varphi$ to the exact
impurity location in each section of length $a$, as discussed
earlier.

\vskip .1in
\parindent 0cm
{\bf III. 4 Scattering through a periodic array of delta scatterers}
\parindent 1cm

The scattering matrix elements for 1D periodic system (or superlattice) can be
calculated exactly \cite{GAAK88,Gas89,GGJRO97}. The transmission amplitude is found to be
\begin{equation}
t_N = {\frac{e^{i(N-1)ka}}{D_{N}}}\label{t},
\end{equation}
and the reflection amplitude is given by
\begin{equation}
r_N =-i\frac{k_0}{k} {\frac{e^{i(N-1)ka}}{D_{N}}}{\frac{\sin({N\beta a})}{\sin ({\beta a})}}, \label{R-}
\end{equation}
where
\begin{equation}
D_N = e^{iNka}\left\{\cos(N\beta a)+i{\it
Im}\left[e^{-ika}(1+i \frac{k_0}{k})\right]\frac{\sin(N\beta a)}{\sin(\beta a)}\right\}
\end{equation}
and $k= \frac{1}{\hbar} \sqrt{2 m^* E}$,
$k_0 = m^* V_I / {\hbar}^2 $, $a$ is the distance between adjacent scatterers, and
$\beta$ is the quasi momentum. It is the solution of the transcendental equation:
\begin{equation}
\cos(\beta a) = \cos(ka)+\frac{k_0}{k} sin(ka).
\end{equation}

Using Eqs.(8-10), it can be shown that
$<S_x>=0$ and
\begin{equation}
\langle S_y\rangle = -{\hbar}{\it Re}({r^*t})=-
\frac{\hbar}{1+ (\frac{k_0}{k})^2 \frac{\sin^2(N\beta a)}{\sin^2(\beta a)}}\frac{\sin(N\beta a)}{
\sin(\beta a)} \frac{k_0}{k}, \label{sy}
\end{equation}
and
\begin{equation}
\langle S_z\rangle=\frac{\hbar}{2}\left( 2|t|^2-1\right)
=\frac{\hbar}{2}\left[
\frac{2}{1+(\frac{k_0}{k})^2 \frac{\sin^2(N\beta a)}{\sin^2(\beta a)}}-1\right]. \label{sz}
\end{equation}

In the case of $N=1$, we get back Eqs. (35) and (36) of section III.1.
In the case of $N$ delta scatterers, incident energies for which
\begin{equation}
\frac{\sin(N\beta a)}{\sin(\beta a)}=0,
\end{equation}
correspond to points of unity transmission which occur at values of the
quasi-momemtum in the first Brillouin zone
\begin{equation}
{\beta}_n a=\frac{\pi n}{N}
\end{equation}
with $(n=1,... N-1)$.

At these values,
$<S_x>= <S_y>=0$ and $<S_z>= \frac{ \hbar}{2}$.

\vskip .1in
\parindent 0cm
{\bf III.5 Transport through random arrays of delta scatterers}
\vskip .1in
\parindent 0cm
The analysis of the previous section was extended to a large number of
delta scatterers of strength
$V_I\delta\left(x-(x_0^i+(i-1)a)\right)$, where $V_I$ is selected to
be 0.3 $eV \AA$ and $x_0^i$ is the location of the $i^{th}$ impurity
located in the interval $\left[(i-1)a,ia\right]$
Each impurity location is generated using a uniform random number in each
interval.  The length of each subsection is set equal to 237 $\AA$
and the wavevector of the incident electron
$k=\frac{\sqrt{2m^\star{E}}}{\hbar}$, is selected such that
$ka=\pi$, for an incident energy $E$ of 10 meV and
$m^\star = 0.067 m_0$, the electron effective mass in GaAs.

\begin{figure}[h]
\centerline{
{\includegraphics[height=6.5in,angle=270]{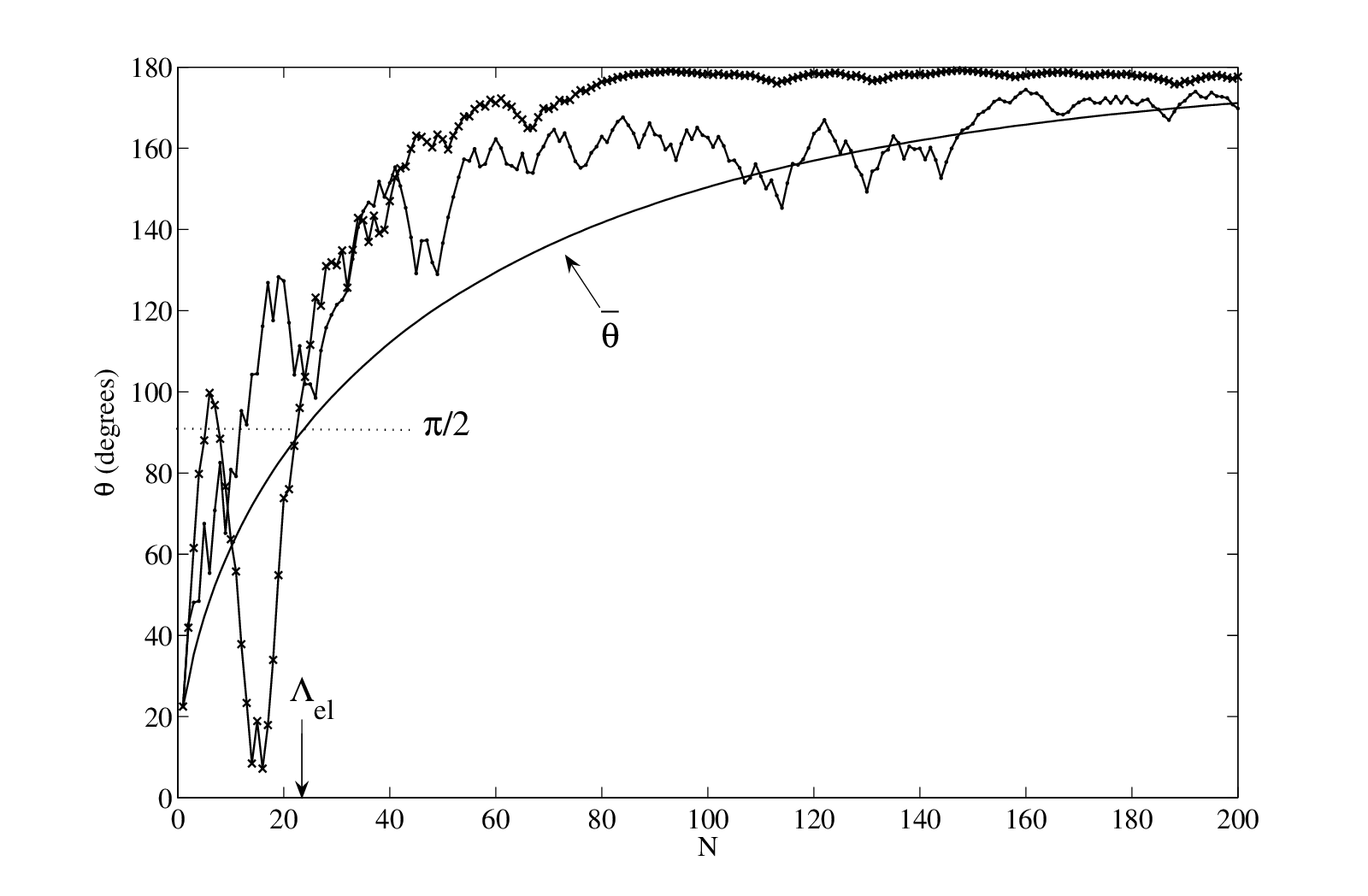}}}
\caption{J. Wan et al}
\label{wanfig9}
\end{figure}

\begin{figure}[h]
\centerline{
{\includegraphics[height=6.5in,angle=90]{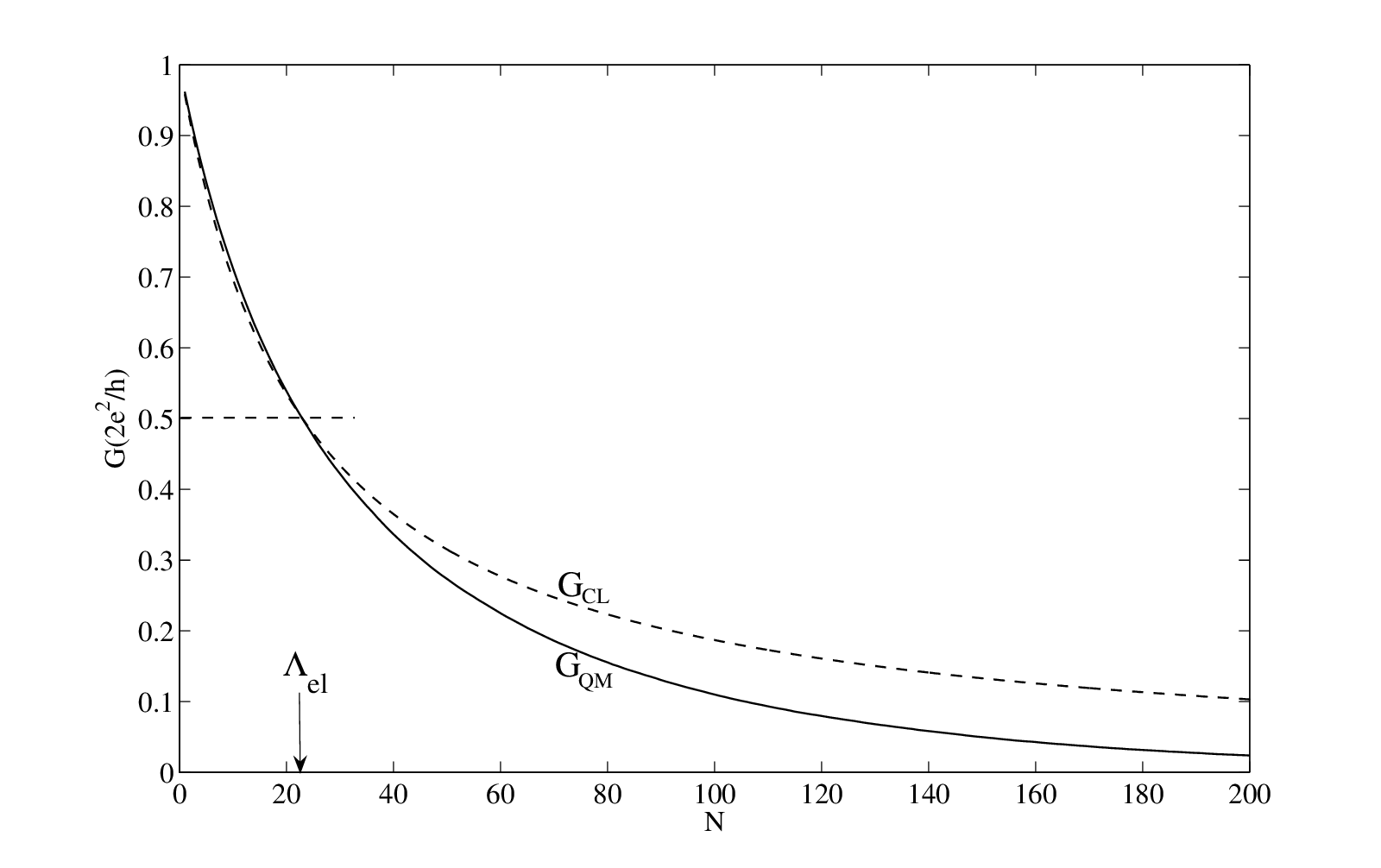}}}
\caption{J. Wan et al}
\label{wanfig10}
\end{figure}

\parindent 1cm
Figure~\ref{wanfig9} is a plot of the phase angle $\theta$ of the spinor
$|\psi^l(OUT)\rangle$ versus the number of subsections ($N$) crossed. The
two top curves are $\theta$ versus $N$ for two specific impurity
configurations. The curves show regions where $\theta$ decreases as
$N$ increases which corresponds to an increase in the conductance of the
array. This decrease in $\theta$ as $N$ increases is quite pronounced
for one of the two impurity configurations, for $N<20$.
A plot of the average value of $\theta$ over an ensemble of $10^5$ samples is
shown as the curve labeled $\overline{\theta}$ in Figure~\ref{wanfig9}. The quantity
$\overline{\theta} = \pi/2$ for $N\approx23$.  This corresponds to a conductance of $e^2/h$,
as shown in Figure~\ref{wanfig10}, and to an elastic mean
free path equal to $23\times 237{\AA} \sim 0.55 \mu m$.

\newpage
\vskip .2in
\begin{center}
\bf{IV. Conclusions}
\end{center}

The effective spin concept examined in this paper offers an alternative
description of phase coherent charge transport through  mesoscopic systems
in terms familiar to researchers in the field of spintronics and quantum
computing.  As illustrated in this paper, the effective spin formalism provides a
pedagogical approach to simple scattering problems and also to
the phenomenon of localization in
random arrays of elastic scatterers.

In the past, the effective spin concept has been used to
describe the spatial correlations between reflection and
transmission amplitudes of polarized photon beams from a combination
of beam splitters, mirrors, and interferometers \cite{holbrow,pittman,cochrane}.
More recently, the effective spin concept has been used to examine the critical
problem of entanglement between channels associated with propagating modes in mesoscopic systems,
as reported in recent experiments by Neder et al. \cite{neder} and their theoretical
interpretation by Samuelson et al. \cite{samuelson}.

\newpage
\parindent 0cm
\vskip .2in
\begin{center}
{\bf Figure Captions}
\end{center}
\vskip .1in
{\bf Figure 1:} The tunneling problem and its quantum
computing gate equivalent. The scattering matrix associated to a device
relates the incoming ($a^+$, $b^-$) to the outgoing ($a^-$, $b^+$)
wave amplitudes. It can be viewed as the matrix representing the rotation
of a qubit from the initial state $|\psi(IN)\rangle$ to the final state
$|\psi(OUT)\rangle$.

\vskip .1in
{\bf Figure 2:} Bloch sphere representation of the effective spin
(qubit) $|\psi(OUT)\rangle$. The radius of the sphere is equal to 1.

\vskip .1in
{\bf Figure 3:} Transmission (T = $|t|^2$) and reflection
(R = $|t|^2$) coefficients versus reduced wavevector $\widetilde{k}$ of
electron incident on a single delta scatterer. The expressions for $|t|$
and $|r|$ are given by Equation (51) and (52), respectively.

\vskip .1in
{\bf Figure 4:} Plot of the phase angles
$(\varphi,\theta)$ associated to the spinor $|\psi(OUT)\rangle$
describing tunneling through a resonant tunneling structure as a
function of the reduced wavevector $\widetilde{k}=k/k_0$, where
$k=\frac{1}{\hbar}\sqrt{2m^\star{E}}$, $E$ is the kinetic energy of
the incident electron in the contact; $k_0=m^\star{V_I}/\hbar^2$,
and $V_I$ is the strength of the delta scatterer. The two delta
scatterers are separated by $50{\AA}$ and have a strength
$V_I=0.3eV{\AA}$. $\Delta \widetilde{k}$ is the minimum wavevector (in reduced units) needed to
realize a spin flip from the south to north poles on the Bloch
sphere. The zeroes in $\theta$ are the locations of the energy
resonances.

\vskip 0.1in
{\bf Figure 5:} Transmission (T) and reflection
(R) coefficients versus reduced wavevector $\widetilde{k}$ of the incident electron for
a superlattice modeled as five delta scatterers of $V_I$ separated
by a distance $a$ ($V_I=0.3eV{\AA}$ and $a = 50{\AA}$). The curves
labeled "1" and "2" are T versus $\widetilde{k}$ for two imperfect
superlattices, i.e., for two arrays of 5 delta scatterers whose positions
are selected uniformly over each interval of length $50{\AA}$.

\vskip 0.1in
{\bf Figure 6:} Reduced wavevector dependence of the
phase angle $\left(\varphi,\theta\right)$ associated to the spinor
$|\psi(OUT)\rangle$ describing tunneling across an array of five
delta scatterers separated by $50{\AA}$ and with a scattering
strength $V_I=0.3eV{\AA}$. The zeroes in $\theta$ are where the
transmission through the superlattice reaches unity.
Also shown as dashed lines are the angles ($\varphi, \theta $) through
two random arrays of elastic scatterers.

\vskip 0.1in
{\bf Figure 7:} Evolution of the angle $\theta$
for the spinor $|\psi(OUT)\rangle$ on the Bloch sphere
as a function of sample length for
two different arrays of elastic scatterers (two top curves).
The smoother curve represents the average of $\theta$ calculated over an average of $10^5$ arrays
with the locations of each individual scatterer varied uniformly
across each subsection of the array.  The elastic mean free path ${\Lambda}_{el}$ (in units of subsections crossed)
is where $ \bar{ {\theta} }$ = $\frac{\pi}{2} $.

\vskip .1in
{\bf Figure 8:} Plot of the average
over an ensemble of $10^5$ impurity configurations
of the conductance as a function of the number of impurities
crossed in the sample. Also, shown is the value of the classical
conductance calculated neglecting the effects of multiple reflections
between scatterers. The elastic mean free path ${\Lambda}_{el}$ (in units of subsections crossed)
is where the Landauer conductance reaches a value of $e^2 /h$.

\newpage
\
\vskip 1in

\newpage
\
\vskip 1in

\newpage
\
\vskip 1in

\newpage
\
\vskip 1in

\newpage
\
\vskip 1in

\newpage
\
\vskip 1in

\newpage
\
\vskip 1in

\newpage
\
\vskip 1in


\begin{thebibliography}{99}
\bibitem{density}
Using the unitary property of the scattering matrix, it can be
easily checked that the trace of the matrix $\rho$ is unity and
$\rho$ satisfies the following properties, $\rho^\dag=\rho$,
$\rho^2=\rho$, and $Tr[\rho^2]=Tr[\rho]=1$, which are all characteristic of the
density matrix associated with a pure state \cite{nielsen}.
\end{thebibliography}
\end{document}